\documentclass[a4paper,12pt]{article}
\usepackage[latin1]{inputenc}
\usepackage{amsfonts,amsthm,amsmath,amssymb,hyperref}

\newcommand{\eqn}[1]{(\ref{#1})}
\def\Tr{{\rm Tr}}

\def\d{\partial}

\def\dag{\dagger}
\newcommand{\be}{\begin{equation}}
\newcommand{\bea}{\begin{eqnarray}}

\newcommand{\ee}{\end{equation}}
\newcommand{\eea}{\end{eqnarray}}

\newcommand{\nn}{\nonumber}

\def\a{\alpha}
\def\b{\beta}

\def\de{\delta}
\def\g{\gamma}
\def\G{\Gamma}
\def\e{\epsilon}

\def\m{\mu}

\def\n{\nu}

\def\o{\omega}

\def\k{\kappa}

\def\s{\sigma}

\def\l{\lambda}

\def\p{\psi}

\def\t{\theta}

\newcommand{\cK}{\mathcal K}
\newcommand{\cL}{\mathcal L}
\newcommand{\cM}{\mathcal M}
\newcommand{\cN}{\mathcal N}

\newcommand{\cP}{\mathcal P}
\newcommand{\cQ}{\mathcal Q}

\newcommand{\cW}{\mathcal W}

\newcommand{\cZ}{\mathcal Z}

\def\hA{\hat{A}}

\title{Boundary Conditions for Interacting Membranes}

\begin{document}

\begin{titlepage}

\begin{center}

\hfill{QMUL-PH-09-29} \\
\hfill{MIFP-09-51} 

\vskip 1cm

{{\Large \bf Boundary Conditions for Interacting Membranes}} \\

\vskip 1.25cm {David S. Berman$^\dag$\footnote{D.S.Berman@qmul.ac.uk}, Malcolm J. Perry$^\ddag$\footnote{M.J.Perry@damtp.cam.ac.uk}, Ergin Sezgin$^\sharp$\footnote{sezgin@tamu.edu}
 and
Daniel C. Thompson$^\dag$\footnote{D.C.Thompson@qmul.ac.uk}}
\\
{\vskip 0.2cm
$^\dag$ Queen Mary University of London,\\
Department of Physics,\\
Mile End Road, London, E1 4NS, England\\
}
{\vskip 0.2cm
$^\ddag$DAMTP, Centre for Mathematical Science,\\
University of Cambridge,\\
Wilberforce Road, CB3 0WA, England\\
}
{\vskip 0.2cm
$^\sharp$George P. and Cynthia W. Mitchell Institute for
Fundamental Physics and Astronomy,\\
Texas A\&M University, College Station, TX 77843-4242, USA}
\end{center}
\vskip 1 cm

\begin{abstract}
\baselineskip=18pt\
We investigate supersymmetric boundary conditions in both the Bagger-Lambert and the ABJM theories of interacting membranes. We find boundary conditions associated to the fivebrane, the ninebrane and the M-theory wave.  For the ABJM theory we are able to understand the enhancement of supersymmetry to produce the (4,4) supersymmetry of the self-dual string.  We also include supersymmetric boundary conditions on the gauge fields that cancel the classical gauge anomaly of the Chern-Simons terms. 
\end{abstract}

\end{titlepage}

\pagestyle{plain}

\baselineskip=19pt
\tableofcontents

\section{Introduction}

 In M-theory the simplest extended object that preserves half of the thirty-two
supersymmetries is the membrane. It is known that a single membrane
may have a boundary, and the resultant open membrane theory will preserve one quarter
of the supersymmetry provided the membrane ends on a fivebrane (for a review and further references to the M-theory literature see \cite{Berman:2007bv}).  The study of single membranes with
boundaries has been undertaken in a variety
of works \cite{Townsend:1995af,Strominger:1995ac,Chu:1997iw,Chu:1998pb,Horava:1995qa,Horava:1996ma,Cederwall:1997hg}.

Through the pioneering work of Bagger and Lambert \cite{Bagger:2006sk,Bagger:2007jr,Bagger:2007vi}  
(and also Gustavsson \cite{Gustavsson:2007vu}) and subsequently Aharony, Bergman, Jafferis and
Maldacena \cite{Aharony:2008ug}, a supersymmetric theory of interacting membranes has been developed. Immediately,
one may ask what can we learn about the theory of interacting open
membranes or, more simply, what is the theory of Bagger, Lambert or ABJM
when the membrane has a boundary? The beginnings of this study have
been undertaken by a variety of authors \cite{Berman:2008be,Berman:2009kj,Nastase:2009ny,Ho:2008ve,Ho:2008nn,Chu:2009ms,Terashima:2008sy,Terashima:2009fy,Hanaki:2008cu}. In this paper we will carry out a systematic study of possible supersymmetric boundary conditions 
and their interpretations. 

Throughout the paper we will follow very closely the
work of Gaiotto and Witten \cite{Gaiotto:2008sa} who carried out a study of boundary conditions in 
${\mathcal{N}}=4$ Yang-Mills theory in four dimensions or
equivalently, interacting D3-branes with boundaries. We will find some similarities to that study and some notable
differences for example,  instead of the Nahm equation  \cite{Nahm:1979yw} we have the Basu-Harvey equation \cite{Basu:2004ed}.  The Basu-Harvey equation was, in fact, developed to describe how membranes end on fivebranes through the analogy with the appearance of the Nahm equation \cite{Nahm:1979yw} in the D1-D3 system  \cite{Diaconescu:1996rk}. There the Nahm equation is the BPS equation of the D1 string. In the work of Gaiotto and Witten \cite{Gaiotto:2008sa}, the Nahm equation also appears as a boundary condition for describing $D3$ branes ending on a $D5$. It is therefore not surprising that we have the Basu-Harvey equation appearing in the context of membrane boundary conditions. 

The ABJM model has only manifest ${\cal N}=6$ supersymmetry. Thus a
application of the techniques to find the boundary conditions
in this case produces the result that the supersymmetry on the
boundary is chiral with four chiral and two antichiral fermions. 
It is thought that monopole operators are crucial to the supersymmetry enhancement of the ABJM model to the full ${\cal N}=8$.  We consider how this situation is altered by the boundary and indeed discover the missing two antichiral
boundary fermions which give the expected ${\cal N}=(4,4)$ supersymmetry of the
self-dual string \cite{Howe:1997ue}. We are also able to give a spacetime interpretation of the world sheet
boundary supersymmetry in terms of orbifolding spacetime.

In addition to the expected $M2-M5$ configuration described above we also find boundary conditions corresponding to $M2-M9$, $M2-M5-M5$ and $M2-MW$. These are all the expected quarter BPS configurations of the membrane \cite{Gauntlett:1997cv,Gibbons:1998hm}. 

The structure of this paper is as follows.  In the remainder of this section we provide some details as to the method we use to obtain supersymmetric boundary conditions.   We also outline some of the spacetime considerations for open branes, their endings on other objects and their preserved supersymmetry.   In section 2 we investigate the Bagger-Lambert theory of multiple membranes.  After introducing the theory in section 2.1 we go onto calculate the boundary conditions in section 2.2.   In section 3 we turn our attention to the ABJM model and also consider the boundary conditions on the gauge fields and the enhancement of supersymmetry.

\subsection{General approach to supersymmetric boundary conditions}

In determining the Euler-Lagrange equations of a Lagrangian field theory one encounters the following term
\be
\label{surfaceterm}
 \int_{\cM} d^m x\,  \d_\m \left( \frac{\de {\cL}}{\de \d_\m\Phi} \de \Phi \right) \ ,  
\ee
which can be written as a surface integral. In theories that are at most quadratic in derivatives this is the only contribution that remains when an action is varied and its Euler-Lagrange equations are used.   When the manifold $\cM$ is non-compact one typically assumes that fields vanish at infinity setting this term to zero.  When $\cM$ has a boundary one must, of course, specify boundary conditions that ensure the above surface term vanishes.

It is interesting to think about the effects of a boundary on supersymmetry.  The boundary breaks  translation invariance and so necessarily must break some supersymmetry. Although a generic boundary condition will completely break supersymmetry, one can ask which boundary conditions preserve as much of the remaining supersymmetry as possible. 

Such boundary conditions can be found by demanding that the component of supercurrent normal to the boundary, evaluated at the boundary, vanishes.   To see why this is the case note that a global supersymmetry variation of the action yields a term
\bea
\label{susysurfaceterm}
\de_{susy} S = \int d^m x\, \d_\m \cK^\m\,, 
\eea
which again can be written as a surface integral.  To preserve supersymmetry we must choose boundary conditions that ensure both $(\ref{susysurfaceterm})$ and $(\ref{surfaceterm})$ vanish.  
For concreteness, assume that $\cM$ is three dimensional with coordinates $\{ x^0,x^1,x^2 \}$ and the boundary is located at $x^2=0$.  Then the condition for   $(\ref{susysurfaceterm})$ to vanish is that $\left. n\cdot \cK \right|_{\d \cM}= 0$ where $n$ is a vector normal to the boundary and so in our case we require that the second component $\left. \cK^2 \right|_{\d \cM} =0$.  Consider the component of the supercurrent normal to the boundary, evaluated at the boundary, which is given by
\bea
\left. J^2 \right|_{\d \cM}  =\left.  \frac{\de \cL}{\de \d_2 \Phi} \delta \Phi \right|_{\d \cM}-\left. \cK^2 \right|_{\d \cM}\  .
\eea
 Since the boundary conditions must ensure $(\ref{surfaceterm})$ equals zero the first term in the right hand side of the above must also equal zero and hence
 \bea
 \left. J^2 \right|_{\d \cM}  =-\left. \cK^2 \right|_{\d \cM}\  . 
 \eea
Therefore boundary conditions for which
 \be
\left. J^2 \right|_{\d \cM}=0  \label{fbc}
\ee
 imply that  $\left. \cK^2 \right|_{\d \cM} =0$  and hence that supersymmetry may be preserved.  The argument described above was used by Gaiotto and Witten \cite{Gaiotto:2008sa} to classify the half-supersymmetric boundary conditions of ${\cal N} = 4$ super Yang--Mills theory in four dimensions.   In what follows we stick closely to this method.   
 
In this work we will restrict our attention to the case of a semi-infinite membrane so that we need only consider the boundary at $x^2=0$.   We will in general only be considering flat membranes with trivial topology.

\subsection{Open Branes and supersymmetry}
Let us consider eleven dimensional supergravity with Lorentz invariance $SO(1,10)$.  One of the known solutions to this theory is the $1+2$ dimensional extended object known as the membrane (or M2).   The presence of membranes extended along $\{x^0,x^1,x^2 \}$ breaks the Lorentz group down to $SO(1,2) \times SO(8)$ and this solution is half BPS; of the thirty-two components of the supersymmetry only sixteen remain.  Therefore, from a Goldstone mode analysis of the broken symmetry, on the world volume one expects to have sixteen supersymmetries and the transverse $SO(8)$ realized as the R-symmetry group.   The thirty-two component Majorana spinor of $SO(1,10)$ decomposes into a sixteen of $SO(1,2) \times SO(8)$ which obeys  $\G^{012}\Psi = -\Psi$.   Since we have that\footnote{In this paper we denote the eleventh dimension by the natural symbol $\natural$.}   $\G^{0123456789\natural}=1$ the condition $\G^{012}\Psi = -\Psi$ actually determines the $SO(8)$ representation of the fermion.    Then the world volume fermion field $\Psi$ transforms in the $2\otimes 8_c$.   On the other hand the supersymmetry parameter is a spinor obeying $\G^{012}\e = \e$ and hence transforming in the $2\otimes 8_s$.

One can further consider the reduction to the $1/4$ BPS sector
corresponding to an M2-M5 system. Then one has
\bea
\label{symansatz}
SO(1,2) \times SO(8) \rightarrow SO(1,1) \times SO(4) \times SO(4)
\eea
where the $SO(1,1)$ corresponds to the symmetry of the boundary
string, one $SO(4)$ is the remaining symmetry of the space in the
fivebrane transverse to
the string and the remaining $SO(4)$ is the symmetry of the space transverse to both the
membrane and fivebrane.

We will consider the spin cover $Spin(4)\cong SU(2)\times SU(2)$   of the $SO(4)$ factors.
An example of this of decomposition is  
\bea
16 \rightarrow (+, 1,2, 2,1)\oplus (-,2,1,1,2)\,, 
\eea
and indeed this example is exactly what one finds for the representations of fields in the self-dual string which is the ending of a membrane on a fivebrane \cite{Berman:2004ew}. At this stage there is no reason to believe this is the only admissible decomposition and there may be other allowed representations.  

Another interesting reduction to consider is
\bea
\label{symansatz2}
SO(1,2) \times SO(8) \rightarrow SO(1,1) \times SO(8)
\eea
which is relevant to the $1/4$ BPS sector corresponding to M2-M9 and M2-MW(ave) systems.  These two cases can be distinguished by the two possibilities for decomposition $2\times 8_s \rightarrow (+, 8_s)$ and $2\times 8_s \rightarrow (-, 8_s)$.

We will look for supersymmetric boundary conditions in the membrane world volume theory that preserve half of the sixteen supersymmetries.   To do so we shall propose an ansatz for the boundary condition of fermionic fields which, in general, requires half of its components to vanish on the boundary.   Given this boundary condition we then demand that we can make the normal component of the supercurrent vanish for certain choices of the supersymmetry parameter which preserve exactly half of the supersymmetries. 

\section{Bagger-Lambert Boundary Conditions}
\subsection{Bagger-Lambert action, SUSY and Supercurrent}
Although the correct theory describing a single membrane has been known for many years \cite{Bergshoeff:1987cm} it was not until the breakthrough of Bagger and Lambert  \cite{Bagger:2006sk,Bagger:2007jr,Bagger:2007vi}   that the full theory of interacting membranes began to be uncovered.   In their approach, Bagger and Lambert suggested that the fields describing multiple membranes do not take values in a Lie algebra (as is the case for the fields describing multiple interacting D-branes) but rather in a (Lie)-Three-algebra.  In this construction the traditional Lie-bracket is replaced with an antisymmetric triple bracket.     

The world volume fields are eight scalars $X^I_a$, a gauge field
$A_{\mu\, ab}$ and fermions $\Psi_a$.  The fermion is a  Majorana
spinor of $SO(1,10)$ restricted by the  projection $\G^{012}\Psi = -
\Psi$. The world volume coordinates  are $\{ x^\mu \} $ and transverse
coordinates $\{ x^I \}$.  The lowercase Roman indices correspond to
the Lie-Three-Algebra, ${\cal A}$, in which fields take their values,
more exactly we can write $X^I = X^I_a T^a$ where $T^a$ are generators
of the Three-Algebra and $a=1\dots N$ where $N$ is the dimension of
the algebra ${\cal A}$.  Structure constants are defined by the triple
bracket as $[T^a, T^b , T^c]= f^{abc}_{\phantom{ abc} d}T^d$. Algebra
indices may be raised or lowered by an inner product which we take to
be $\delta_{ab}$.  The requirement that the bracket is compatible with
the inner product implies that $f^{abcd}= f^{abc}_{\phantom{ abc}
  e}\delta^{ed}$ is totally antisymmetric. The gauge field, which
carries two three-algebra indices should be thought of as living in
the space of linear maps from the three-algebra to itself.
 
The joint requirements of the total antisymmetry of the triple bracket and the positivity of an inner product on the algebra are very constraining.  In fact, there is an essentially unique positive-definite Three-Lie algebra known as ${\cal A}_4$ \cite{Nagy,Papadopoulos:2008sk,Gauntlett:2008uf}.  In this case the structure constant $f^{abcd} = \e^{abcd}$ is the invariant form on $SO(4)$.  Despite much work in generalising the algebraic structure \cite{Benvenuti:2008bt,Gomis:2008uv,Bandres:2008kj}, the ${\cal A}_4$ model remains the only example of a unitary three-dimensional interacting theory with manifest maximal supersymmetry.  Despite this, we find it very useful to study the Bagger-Lambert theory since the manifest ${\cal N} = 8$ supersymmetry allows for easier interpretations of the brane configurations that our boundary conditions describe.  

The Lagrangian is given by
\bea
\cL &=& -\frac{1}{2}D_\m X^{aI} D^\m X_a^I + i \bar{\Psi}^a \Gamma^\mu D_\mu \Psi_a + i \bar{\Psi}_a \Gamma_{IJ} X^I_bX^J_c \Psi_d f^{abcd} \nn \\
&&-V_{b} +\frac{1}{2}\e^{\m\n\l} \left( f^{abcd} A_{\m ab} \d_\n A_{\l cd} + \frac{2}{3}f^{abcg}f^{def}_{\phantom{abc}g} A_{\m ab}A_{\n cd} A_{\l ef} \right)\, . 
\eea
The bosonic potential $V_{b}$ is sextic and is essentially given by the square of the three bracket as
\bea
\label{Vbose}
V_b = \frac{1}{12} f^{abcd} X_a^I X_b^J X_c^K f^{efg}_{\phantom{efg}d} X_e^I X_f^J X_g^K    . 
\eea
The supersymmetry rules are
\bea
\de X^I_a &=& i\bar{\e} \G^I \Psi_a \, , \nn \\
\de \Psi_a& =& D_\m X_a^I \G^\m \G^I \e - \frac{1}{6} X^I_b X^J_c X^K_d f^{bcd}_{\phantom{bcd}a} \G^{IJK} \e\, , \nn\\
\de \tilde{A}_{\mu\, a}^b &=& i \bar{\e}\G_\m \G_I X^I_c \Psi_d
f^{bcd}_{\phantom{bcd} a} \, ,  
\eea 
with $\G^{012}\epsilon = \epsilon$ and $\tilde{A}_{\mu\, a}^b = f^{cdb}_{\phantom{cdb}a} A_{\mu cd}$. The supercurrent is given by \cite{Passerini:2008qt}:
\bea
J^\m =- \bar{\e} D_\n X^I_a \G^\n \G^I \G^\mu \Psi^a - \frac{1}{6}
\bar{\e} X^I_a X^J_b X^K_c f^{abcd}\G^{IJK}\G^\m \Psi_d\, \ . \label{J}
\eea

\subsection{BL Boundary conditions}
One may now simply insert (\ref{J}) into (\ref{fbc}) and evaluate the resultant expressions to determine the supersymmetric boundary conditions 
\bea
\label{blboundarycondition}
0=\left(- \bar{\e}D_\n X^I_a \G^\n \G^I \G^2 \Psi^a - \frac{1}{6} \bar{\e} X^I_a X^J_b X^K_c f^{abcd}\G^{IJK}\G^2 \Psi_d\right) \left. \right|_{\d \cM} \,.
\eea 
To solve this equation we adopt an ansatz for the preserved symmetry of the solution, this is given by the $SO(4)\times SO(4)$ structure of (\ref{symansatz}).   We thus decompose the scalar fields into two $4 {}s$ of $SO(4)$ by writing $X^A = \{X^3,X^4,X^5,X^6 \}$ and $Y^P = \{ X^7, X^8,X^9,X^\natural\}$.  We adopt the following notation: hatted Greek run over $\{0,1\}$, early capital Roman run over $\{ 3,4,5,6\}$ and late capital Roman over $\{7,8,9,\natural\}$.  It is also convenient to define $\Gamma_2 \Psi_a = \tilde{\Psi}_a$; nothing has been lost in doing so. 

Then ($\ref{blboundarycondition}$) becomes
\bea
0&=& -\bar{\e}D_{\hat{\n}} X^A_a \G^{\hat{\n}} \G^A \tilde{\Psi}^a  \nn \\
&& - \bar{\e}D_{\hat{\n}}  Y^P_a \G^{\hat{\n}}\G^P \tilde{\Psi}^a  \nn \\
&&  - \bar{\e}\left( D_2 Y^P_a \G^2 \G^P \de^{da} + \frac{1}{6} Y_a^PY_b^QY_c^R f^{abcd} \G^{PQR}\right) \tilde{\Psi}_d  \nn \\
&& - \bar{\e}\left( D_2X^A_a \G^2 \G^A \de^{da} + \frac{1}{6} X_a^AX_b^BX_c^C f^{abcd} \G^{ABC}\right) \tilde{\Psi}_d  \nn \\
&&  - \bar{\e}\left( \frac{1}{2} X_a^AX_b^BY_c^P f^{abcd} \G^{ABP}\right) \tilde{\Psi}_d  \nn \\
&&  - \bar{\e}\left( \frac{1}{2} X_a^AY_b^P Y_c^Q f^{abcd} \G^{APQ}\right) \tilde{\Psi}_d 
\eea
where $\hat{\mu} = 0,1$.  
We have grouped terms together according to their Lorentz structure. In general there is no reason why this equation can't be solved by canceling across these different groupings.  However, we are only interested in solutions which preserve the $SO(1,1) \times SO(4)\times SO(4)$ symmetry and hence demand that each of the above lines be zero separately:  
  \bea
0&=& \bar{\e}D_{\hat{\n}} X^A_a \G^{\hat{\n}} \G^A \tilde{\Psi}^a \label{eq1}  \\
0&=&  \bar{\e}D_{\hat{\n}}  Y^P_a \G^{\hat{\n}}\G^P \tilde{\Psi}^a  \label{eq2}\\
0&=&  \bar{\e}\left( D_2 Y^P_a \G^2 \G^P \de^{da} + \frac{1}{6} Y_a^PY_b^QY_c^R f^{abcd} \G^{PQR}\right) \tilde{\Psi}_d  \label{eq3} \\
0&=&\bar{\e}\left( D_2X^A_a \G^2 \G^A \de^{da} + \frac{1}{6} X_a^AX_b^BX_c^C f^{abcd} \G^{ABC}\right) \tilde{\Psi}_d \label{eq4} \\
0&=&  \bar{\e}\left( \frac{1}{2} X_a^AX_b^BY_c^P f^{abcd} \G^{ABP}\right) \tilde{\Psi}_d  \label{eq5} \\
0&=&  \bar{\e}\left( \frac{1}{2} X_a^AY_b^P Y_c^Q f^{abcd} \G^{APQ}\right) \tilde{\Psi}_d \label{eq6}
\eea

The next step is to solve the above equations.  We do this on a case by case basis according to the number of scalars obeying Dirichlet boundary conditions.  We propose ansatz for the fermionic boundary conditions and search for the preserved supersymmetry.

\subsubsection{Case 1. Half Dirichlet (M5)}
We will group our solutions according to how many of the scalar fields obey Dirichlet conditions.   First of all we assume that exactly half of the scalars (which we will assume to be the $Y^P$) obey Dirichlet conditions $D_{\hat{\mu}} Y^P = 0$.  For simplicity we will also assume that this is solved with $Y^P= 0$ on the boundary although we note that there could be some interesting scenarios obtained by relaxing this assumption. 

Given this boundary condition on the scalars we are left with only three equations to satisfy:
\bea
\label{eqA} 0&=& \bar{\e} D_2 Y^P \G^2 \G^P \tilde{\Psi} \ ,\\
\label{eqB} 0&=& \bar{\e} D_{\hat{\n}}X^A \G^{\hat{\n}} \G^A \tilde{\Psi} \ ,  \\
\label{eqC} 0&=& \bar{\e} \left(D_2 X_a^A\delta^{ad} \G^2 \G^A + \frac{1}{6}f^{abcd} X^A_aX^B_bX^C_c \G^{ABC} \right) \tilde{\Psi}_d \ .
\eea
We must solve \eqn{eqA} and do not wish to set $D_2 Y = 0$, (Y can't be simultaneously D and N).  We propose an ansatz for the fermion
\bea
\frac{1}{2}(1- \G^{013456} ) \tilde{\Psi}=\cQ_- \tilde{\Psi}     =0 \ ,
\eea   
in which we introduce a projector $\cQ_\pm = \frac{1}{2}(1 \pm \G^{013456} )$.   
Then \eqn{eqA} implies the supersymmetry must obey
\bea
\label{m5susy}
(1-\G^{013456} ) \e=\cQ_-\e     =0\,.
\eea 
These choices automatically ensure that ($\ref{eqB}$) holds so all that remains is to choose boundary conditions on the remaining scalars to satisfy ($\ref{eqC}$).  We wish to remove the inhomogeneous gamma matrix structure from ($\ref{eqC}$) which we can do by means of the identity $\Gamma^A = \frac{1}{6} \e^{ABCD} \G^{BCD}\G^{3456}$.  Then after employing the projector conditions on the fermions we are left with 
\bea
0&=& \left(D_2X^D_a\e^{DABC}  \de^{da} + X_a^AX_b^BX_c^C f^{abcd} \right)  \bar{\e}\G^{ABC} \tilde{\Psi}_d  \, 
\eea
which is solved having the scalars $X^A$ obey Basu-Harvey type equations 
\bea
\label{BHeqn}
0 = D_2X^A_a + \kappa \e^{ABCD} X^B_b X^C_c X^D_d  f^{bcd}_{\phantom{abc}a}  \, , 
\eea
 with $\kappa={1 \over 6}$ in this choice of conventions.

In view of (\ref{m5susy}) and that $\Gamma^{012}\e = \e$, the preserved supersymmetry is in $(+, 1,2,1,2) \oplus (-,2,1,2,1)$ and the fermion that is not projected out is in $(+,2,1,1,2) \oplus (-,1,2,2,1)$.
 This system represents membranes ending on a five-brane with the following brane picture.
\bea
\begin{array}{c|ccccccccccc} 
& x^0 & x^1 & x^2 &x^3 & x^4 & x^5 & x^6 & x^7 & x^8 & x^9 & x^{11}\\ \hline
M2 & - & - & - & \cdot & \cdot & \cdot & \cdot & \cdot & \cdot & \cdot& \cdot\\
M5 & - & - & \cdot & - & -& - & - &  \cdot& \cdot& \cdot &\cdot
\end{array}
\eea

There is another known M-theory brane configuration which we might also expect to find from the membrane boundary conditions.  This is given by the $M2-M5-M5$ intersection of the following diagram.
\bea
\begin{array}{c|ccccccccccc} 
& x^0 & x^1 & x^2 &x^3 & x^4 & x^5 & x^6 & x^7 & x^8 & x^9 & x^{11}\\ \hline
M2 & - & - & - & \cdot & \cdot & \cdot & \cdot & \cdot & \cdot & \cdot& \cdot\\
M5 & - & - & \cdot & - & - & -& - &\cdot& \cdot&\cdot &\cdot\\
M5 & - & - & \cdot &\cdot&\cdot&\cdot&\cdot & -&-&-&-
\end{array}
\eea
From a spacetime perspective such a configuration is also $\frac{1}{4}$ BPS with the supersymmetry parameter obeying
\bea
\G^{012}\e = \e \ , \\
\G^{013456}\e = \e \ , \\
\G^{01789\natural}\e = \e \ . 
\eea 
The reason that this is $\frac{1}{4}$ BPS rather than $\frac{1}{8}$ BPS as naive counting would dictate is that not all of these projections are independent. In fact the first two, together with the identity $1= \Gamma^{0123456789\natural}$ imply the third.  Of course, these supersymmetry projections are exactly the ones we are using already.  Essentially we are free to add one additional fivebrane with impunity.  We obey the same Dirichlet and Neumann conditions as before including the Basu-Harvey equation (\ref{BHeqn}).

At this point we note that we have not determined the boundary conditions for the gauge fields.  This is because, unlike the ${\cal N}=4$ super Yang-Mills theory studied in \cite{Gaiotto:2008sa}, the gauge field only enters into the supercurrent through covariant derivatives of matter fields.  Hence there is no equation constraining the field strength which would give rise to standard Neumann or Dirichlet boundary conditions for the gauge field. A related point is that the kinetic term for the gauge fields, which is essentially a Chern-Simons term, is not exactly gauge invariant, its variation produces a boundary term which we must include in our analysis.   We will study these issues in more detail in the context of the ABJM theory in the next section. We will see that we can recover gauge invariance by putting boundary conditions, which are non-derivative algebraic constraints, on the $A_0$ and $A_1$ components of the gauge field.  This argument does not fix a boundary condition on $A_2$.  Instead the behaviour of $A_2$ is algebraically determined from that of the scalars through the Basu-Harvey equation.    A further question is that of the closure of the boundary conditions.  Since we will address this in some detail for the ABJM theory we shall not discuss this for the Bagger-Lambert case  to avoid repetition.

\subsubsection{Case 2.  No Dirichlet (M9)}
We now assume none of the scalars obey Dirichlet boundary conditions.  We propose an ansatz for the fermionic boundary conditions motivated by the apparent restoration of $SO(8)$ symmetry exhibited by the scalars:
\bea
\label{fermbcansatz}
\frac{1}{2}(1+ \G^{013456789\natural} ) \tilde{\Psi} = \frac{1}{2} (1+ \G^{2} ) \tilde{\Psi}= \cP_+ \tilde{\Psi} =0 \, ,
\eea
in which $\cP_{\pm} = \frac{1}{2} \left( 1 \pm \G^2 \right) $
First we must solve for \eqn{eq1} and \eqn{eq2} assuming that $D_{\hat{\n}} X \neq 0$ and so require
\bea
0 = \bar{\e} \G^{\hat{\n}} \G^I \tilde{\Psi} , & \, I = 3\dots \natural . 
\eea
We insert the ansatz for fermion boundary condition (\ref{fermbcansatz}) and have 
\bea
0 &=& \bar{\e} \G^{\hat{\n}} \G^I\cP_-\tilde{\Psi} = \bar{\e}\cP_- \G^{\hat{\n}} \G^I\tilde{\Psi}\, ,
\eea
so we conclude that the supersymmetry must obey 
\bea
\cP_+ \e = 0\,.
\eea
One can now read off the representations for the preserved supersymmetry and the fermion boundary condition.  A little care is in order; the projector $\cP_+ = \frac{1}{2}\left( 1 + \G^2 \right)$ does not simply project out $SO(1,1)$ chiralities, it actually picks out the product of $SO(1,1)$ and $SO(8)$ chiralities.  So for the supersymmetry parameter we have $\G^{012}\e = \e$ and $\G^2 \e = -\e$ from which we conclude that $\e$ is in the $(-, 8_s)$  of $SO(1,1) \times SO(8)$.  Likewise for the fermion, the components that are not constrained to zero  by the boundary condition obey $\G^{012}\tilde{\Psi} = -\tilde{\Psi}$ and $\G^2 \tilde{\Psi} = -\tilde{\Psi}$ and are thus in the $(+, 8_c)$ 

Now we must check (\ref{eq5}) and (\ref{eq6}):
\bea
0&=&  \bar{\e}\left( \frac{1}{2} X_a^AX_b^BY_c^P f^{abcd} \G^{ABP}\right) \tilde{\Psi}_d \,  \\
0&=&  \bar{\e}\left( \frac{1}{2} X_a^AY_b^P Y_c^Q f^{abcd} \G^{APQ}\right) \tilde{\Psi}_d \ .
\eea
Since we have that $\G^{IJK} \cP_- = \cP_+ \G^{IJK}$ these equations are not automatically satisfied and we find the constraint
\bea
 X_a^AX_b^BY_c^P f^{abcd} =X_a^AY_b^P Y_c^Q f^{abcd}= 0 
\eea
We also find from  \eqn{eq3} and  \eqn{eq4}:
\bea
\label{M9bceqn1}
0&=& \bar{\e}\left( D_2 Y^P_a  \G^2\G^P \de^{da} + \frac{1}{6} Y_a^PY_b^QY_c^R f^{abcd} \G^{PQR}\right) \tilde{\Psi}_d \, ,   \\
\label{M9bceqn2} 0&=&   \bar{\e} \left( D_2X^A_a \G^2 \G^A \de^{da} + \frac{1}{6} X_a^AX_b^BX_c^C f^{abcd} \G^{ABC}\right) \tilde{\Psi}_d  \, .
\eea
An obvious solution is to demand
\bea
\label{M9sols}
D_2 Y^P_a=D_2 X^A_a=0, & \,   X_a^AX_b^BX_c^C f^{abcd}=Y_a^PY_b^QY_c^R f^{abcd}=0 .
\eea
Then the bosonic boundary conditions can be written in an $SO(8)$ covariant way
\bea
D_2 X_a^I = 0&  X_a^IX_b^JX_c^K f^{abcd}=0\,.  \label{ybc}
\eea

One way to understand the conditions described by equation (\ref{ybc})
is to see that the algebraic constraint is equivalent to demanding
that the potential vanishes (this expression squared gives the
potential in Bagger-Lambert theory (\ref{Vbose})).  It then seems that the boundary condition suggests that  the boundary string lives in the moduli
space of the membrane theory which for Bagger-Lambert theory is given
by \cite{Distler:2008mk,Lambert:2008et} ${\cal M}_{BL} = \frac{\mathbb{R}^8\times\mathbb{R}^8}{D_{2k}}$  where $k$ is the overall level of the theory (previously set to one).  To make this identification complete one needs to be careful about the gauge field.

Given the chiral nature of the supersymmetry and fermion projector and $SO(8)$ structure of these boundary conditions there seems to be a natural interpretation in terms of a membrane ending on a M9 described by the following brane diagram:
\bea
\begin{array}{c|ccccccccccc} 
& x^0 & x^1 & x^2 &x^3 & x^4 & x^5 & x^6 & x^7 & x^8 & x^9 & x^{11}\\ \hline
M2 & - & - & - & \cdot & \cdot & \cdot & \cdot & \cdot & \cdot & \cdot& \cdot\\
M9 & - & - & \cdot & - & - & -& - & -& -&- &-
\end{array}
\eea  

We now ask if there is another way to solve the equations (\ref{M9bceqn1}) and (\ref{M9bceqn2}) other than the $SO(8)$ covariant way described above.   To do so it seems that we need to demand that $\e$ is a eigenstate for $\Gamma^{013456}$ (and hence also $\Gamma^{01789\natural}$) then we have the possible solutions
\bea
D_2 Y^P_a\epsilon^{PQRS}\delta^{ad}  +  Y_a^PY_b^QY_c^R f^{abcd}  = 0 \\
D_2 X^A_a\epsilon^{ABCD}\delta^{ad} -  X_a^B X_b^C X_c^D f^{abcd}  = 0 
\eea
however this option projects out all but $4/32$ components supersymmetry and is thus not a $\frac{1}{2}$ BPS boundary condition.  

One might also try to find boundary conditions for this no Dirichlet case using a different $\frac{1}{2}$ BPS ansatz, for example, by invoking the projectors of the $M2-M5$ case discussed earlier on the supersymmetry parameter.  If one does so, very quickly one finds that the only way to solve the boundary conditions is if the fermion is identically zero.  This does not represent a legitimate choice of boundary conditions.

\subsubsection{Case 3. All Dirichlet (M-Wave)} 
We now assume that all of the scalars are Dirichlet and, for simplicity, that they are zero.  
Then the only equations to satisfy are 
\bea
0&=& \bar{\e}D_{2} X^A_a \G^{2} \G^A \tilde{\Psi}^a  \, ,  \\
0&=&  \bar{\e}D_{2}  Y^P_a \G^{2}\G^P \tilde{\Psi}^a  \, .
\eea
We assume the same ansatz for the fermion as in the preceding case, namely, $(1+\G^2)\tilde{\Psi} = 0$.  We can thus satisfy the above equations by demanding $(1-\G^2)\e = 0$.   This preserves supersymmetry $(+,8_s)$ and has fermions not projected out transforming in $(+, 8_c)$. 

What does this represent? Let us consider reducing the theory down to ten dimensions by performing a double dimensional reduction along the world volume of the brane.  If we did this along the direction in which which the boundary string is extended (that is in the $x^1$ direction) we would be left with an open string with Dirichlet boundary conditions in all eight of its transverse directions.   This has the stringy interpretation of a string ending on a D0 brane.   The M-theory lift of the D0 brane is the M-theory gravitational wave MW.  Hence it seems natural, if slightly unusual, to think of these boundary conditions as corresponding to a membrane ``ending" on a M-wave.

\section{ABJM Boundary Conditions}
The Bagger-Lambert theory studied in the previous section is now known not to be the full theory describing an arbitrary number of interacting membranes.  The algebraic construction is very restrictive which leads to an essentially unique Bagger-Lambert theory which describes the case of two  interacting membranes.   To generalize this construction one must relax some assumptions.  The ABJM theory \cite{Aharony:2008ug} dispenses with the presumed manifest $SO(8)$ R-symmetry and instead displays only manifest $SU(4)$ R-symmetry and hence only ${\cal N}=6$ supersymmetry. In the ABJM approach the algebra is also much more conventional, matter fields (denoted by $Y^A$ for the bosons and $\Psi_A$ for the fermions) are bifundamentals of a $U(N)\times U(N)$ gauge group with Chern-Simons kinetic terms for gauge fields (denoted by $A_{\mu}$ and $\hat{A}_{\mu}$).  This ABJM model is though to properly describe the low energy dynamics of any number of interacting branes whose transverse space is the orbifold $\mathbb{C}^4/\mathbb{Z}_k$ where $k$ is the Chern-Simons level.
 
The gauge group of these theories is not arbitrary, in fact to preserve ${\cal N}=6$ supersymmetry the only other options are $SU(N)\times SU(M)$,  $U(N)\times U(M)$ and $SO(2)\times Sp(N)$ \cite{Aharony:2008ug,Aharony:2008gk,Hosomichi:2008jb}.  Also of note is that the three algebra language of Bagger-Lambert can also be used to describe these sorts of theories by suitably relaxing the form of the three algebra structure constants \cite{Bagger:2008se}.   In what follows we will restrict our attention to the original ABJM  $U(N)\times U(N)$ model and choose not to work in three algebra language.  
\subsection{The ABJM action, supersymmetry and supercurrent}

The ABJM Lagrangian is given by \cite{Benna:2008zy,Bandres:2008ry,Terashima:2008sy}
\bea
 {\cal L} &=& \frac{k}{4\pi} \e^{\m\n\l} \Tr \left( A_\m \d_\n A_\l + \frac{2i}{3} A_\m A_\n A_\l      -    \hA_\m \d_\n \hA_\l - \frac{2i}{3} \hA_\m \hA_\n \hA_\l  \right) \nn \\
 &&\, - \Tr \left( D_\m Y^\dag_A D^\m Y_A  - i \p^{\dag A} \g^\m D_\m \p_A  \right)\nn \\
 &&\, -V_{f} - V_{b}\,, 
\eea
where the sextic bosonic potential is
\bea
V_{b} &=&  \frac{4\pi^2}{3 k^2} \Tr \left( Y^A Y^\dag_A Y^B Y^\dag_B Y^C Y^\dag_C   +Y^\dag_A Y^A Y^\dag_B Y^BY^\dag_C Y^C       \nn \right. \\
 && \qquad \left. +  4 Y^A Y^\dag_B Y^C Y^\dag_A Y^B Y^\dag_C  - 6 Y^A Y^\dag_B Y^B Y^\dag_A Y^C Y^\dag_C \right)
\eea
and the bose fermi interaction terms are 
\bea
V_{f} &=&  \frac{2\pi i}{ k} \Tr   \left(  Y^\dag_A Y^A \p^{\dag B} \p_B    - \p^{\dag B}  Y^A Y^\dag_A \p_B  - 2 Y^\dag_A Y^B \p^{\dag A} \p_B   +2    \p^{\dag B}   Y^A Y^\dag_B\p_A \nn \right. \\
&& \qquad \left.  - \e^{ABCD}  Y^\dag_A \p_B Y^\dag_C \p_D + \e_{ABCD} Y^A \p^{\dag B} Y^C \p^{\dag D}  \right)\,. 
\eea
The matter field $Y^A$ carries an $SU(4)$ index $A$ and transforms in a bifundamental representation of the gauge groups.  The appropriate covariant derivative is thus given by
\be
D_\mu Y^A = \d_\mu Y^A + i A_\mu Y^A - i Y^A \hA_\mu \,. 
\ee
The $\cN=6$ supersymmetry transformations are given by \cite{Terashima:2008sy,Bandres:2008ry}
\bea
\de Y^A &=& i \o^{AB} \p_B \, , \nn \\
\de Y^\dag_A &=& i \p^{\dag B} \o_{AB} \, , \nn \\
\de \p_A &=& \g^\m \o_{AB} D_\m Y^B + N_A \, , \nn \\
\de \p^{\dag A} &=& -D_\m Y^\dag_B  \o^{AB}\g_\mu + N^{\dag A} \, , \nn\\
\de A_\m &=& \frac{2\pi}{k}\left( Y^B \p^{\dag A}\g_\m \o_{AB} + \o^{AB} \g_\m \p_A Y^\dag_B   \right)\, ,  \nn \\
\de \hA_\m &=& \frac{2\pi}{k}\left( \p^{\dag A}Y^B\g_\m \o_{AB} + \o^{AB} \g_\m Y^\dag_B\p_A     \right)\, ,
\eea
where the interaction term in fermion variation is
\bea
N_A =\frac{2\pi}{k}\left( \o_{AB} ( Y^C Y^\dag_C Y^B - Y^B Y^\dag_C Y^C) -  2 \o_{CD}Y^C Y^\dag_A Y^D  \right)\, , \nn \\
N^{\dag A} = \frac{2\pi}{k}\left(  (Y^\dag_B Y^C Y^\dag_C  - Y^\dag_C Y^C Y^\dag_B)\o^{AB} - 2 Y^\dag_D Y^A Y^\dag_C \o^{CD} \right) \, . 
\eea
In these variations the parameter $\o_{AB}$ transforms in the antisymmetric $6$ of $SU(4)$ and caries a $SO(1,2)$ spinorial index which is suppressed\footnote{See appendix for details of SO(1,2) spinor conventions.}.   The important convention to remember is that the spinor contraction involves no additional complex conjugation which is a quite legitimate choice in three dimensions.   Although it seems that there are too many parameters for this to be $\cN=6$ supersymmetry the $\o^{AB}$ and $\o_{AB}$ are not independent; they obey 
\bea
\label{omegaid}
(\o^{AB})_{\a} = -\frac{1}{2} \e^{ABCD} (\o_{CD})_\a = -(\o_{AB})^\star_{\a}  \,. 
\eea
To understand this better it is helpful to recast the supersymmetry in a way which looks more like six copies of $\cN = 1$ supersymmetry.  This is achieved by means of a set of $SO(6)$ gamma matrices $\G^i_{AB}$ with $i=1\dots 6$ which obey
\bea
\G^i \tilde{\G}^{j} + \G^j \tilde{\G}^{ i}= 2 \delta^{ij}\, ,   
\eea
where 
\bea 
\label{tildeG}
\tilde{\Gamma}^{i AB} = -\frac{1}{2} \e^{ABCD} \G^i_{CD} = -(\G^{i}_{AB})^{\star}\,. 
\eea
Then the supersymmetry parameter can be written as $\o_{AB} \equiv \e^i \G^i_{AB}$ where $\e^i$ are six, two-component Majorana spinors.   The reality condition then follows from  (\ref{tildeG}).

The ABJM Lagrangian is invariant under supersymmetry up to a total derivative. Using the Noether procedure one can calculate the supercurrent which is given by \cite{Low:2009kv} as 
\bea
J^\mu = - D_\n Y^\dag_B \o^{AB} \g^\n \g^\m \p_A + N^\dag_A \g^\m \p_A + \p^{\dag A} \g^\m \g^\n \o_{AB} D_\n Y^B + \p^{\dag A} \g^\m N_A\, .
\eea
In the following we reduce notation by introducing a bracket 
\bea
[Y^A, Y^B; Y^\dag_C] \equiv Y^A Y_C^\dag Y^B - Y^B Y^\dag_C Y^A \, .
\eea

\subsection{Boundary Conditions}
We proceed as with the BL case.   However now we must be a little more careful about what symmetry we expect to be preserved.  
The eight transverse scalars are encoded in $Y^A$ which is a $4$ of $SU(4)$ rather than an $8$ of $SO(8)$.  In the BL case we looked for boundary conditions that preserved $SO(1,1) \times SO(4) \times SO(4) \subset SO(1,2) \times SO(8)$.  In the BL theory this causes us to group the scalars into sets of four which share the same boundary conditions.   Although we can not have exactly the same symmetry in the ABJM theory we can still group  the scalars into sets of four sharing the same boundary conditions by looking for b.c. that preserve $SO(1,1) \times SU(2) \times SU(2) \subset SO(1,2) \times SU(4)$ symmetry. 
We thus make the following decomposition for scalars: 
\bea
Y^A =  
\left(\begin{array}{c}  
	X^1 + i X^5 \\
	X^2+i X^6 \\
	X^3 - i X^7 \\
	X^4 - iX^8
\end{array}\right)= (X^a, Y^i) \, ,
\eea
with
\bea
X^a =  
\left(\begin{array}{c}      
	X^1 + i X^5 \\
	X^2+i X^6 \\
	\end{array}\right)\, , 
&Y^i =  
\left(\begin{array}{c}   
	X^3 - i X^7 \\
	X^4 - iX^8
\end{array}\right)\, , 
\eea
The fermion decomposes similarly as $\Psi_A = (\chi_a, \xi_i)$ and the supersymmetry parameters decompose as 
\bea
\o_{AB} =  \left(\begin{array}{cc}      
	\e_{ab} \o & \o_{ai} \\
	-\o_{ai} & \e_{ij} \tilde{\o} \\
	\end{array}\right)\, . 
\eea
Note that the entries of $\o_{AB}$ are complex however the reality condition ensures that we have, in particular, 
\bea
\tilde{\o} = \o^\star \,. 
\eea
We now decompose the super-current along these lines and, as with the BL case, we demand that terms of different Lorentz symmetry in $J^2|_{\partial {\cal M}}$  vanish separately.  This leaves the following equations to be solved by our boundary conditions:
\bea
0 &=& \e_{ab} \tilde{\chi}^{\dag a} \g^2 \o D_2 X^b + \e_{ab}  \tilde{\chi}^{\dag a} \o \k [X^b, X^c; X^\dag_c] +  \e_{cd}  \tilde{\chi}^{\dag a} \o \k [X^c, X^d; X^\dag_a] \, , \\
0 &=& \tilde{\xi}^{\dag i} \g^2 \o_{ai} D_2 X^a +   \tilde{\xi}^{\dag i} \o_{ai} \k [X^a, X^c;  X^\dag_c] \, ,\\
0 &=& \e_{ij} \tilde{\xi}^{\dag i} \g^2 \o^\star D_2 Y^j + \e_{ij} \tilde{\xi}^{\dag i}  \o^\star  \k [Y^j, Y^k ; Y^\dag_k] + \e_{jk} \tilde{\xi}^{\dag i}  \o^\star  \k [Y^j, Y^k ; Y^\dag_i] \, , \\
0 &=&   \tilde{\chi}^{\dag a} \g^2 \o_{ai} D_2 Y^i +  \tilde{\chi}^{\dag a} \o_{ai} \k [Y^i, Y^j ; Y^\dag_j] \, , \\
0 &=&  \e_{ab} \tilde{\chi}^{\dag a} \g^{\hat{\m}} \o D_{\hat{\mu}} X^b  \, , \\
0 &=&    \tilde{\chi}^{\dag a} \g^{\hat{\m}} \o_{ai} D_{\hat{\mu}} Y^i  \, , \\
0 &=& \e_{ij} \tilde{\xi}^{\dag i} \g^{\hat{\m}} \o^\star D_{\hat{\mu}} Y^j \, ,\\
0 &=& \tilde{\xi}^{\dag i} \g^{\hat{\m}} \o_{ai}  D_{\hat{\mu}} X^a \, , \\
0 &=& - \tilde{\xi}^{\dag i}  \o_{ai} \k [ X^a, Y^j ; Y_j^\dag]  +2 \tilde{\xi}^{\dag i}  \o_{aj} \k [ X^a, Y^j ; Y_i^\dag]  \, , \\ 
0 &=&    \tilde{\chi}^{\dag a}  \o_{ai} \k [ Y^i, X^b ; X_b^\dag]  - 2 \tilde{\chi}^{\dag a}  \o_{bi} \k [ Y^i, X^b ; X_a^\dag]  \, , \\
0 &=& \e_{ab}  \tilde{\chi}^{\dag a}  \o [X^b , Y^i; Y^\dag_i] \, , \\ 
0 &=& \e_{ij}\tilde{\xi}^{\dag i}  \o^\star [Y^j, X^c ; X^\dag_c] \, , \\ 
0 &=&   \e_{ab}  \tilde{\xi}^{\dag i}  \o [X^a , X^b; Y^\dag_i] \, , \\
0 &=& \e_{jk}    \tilde{\chi}^{\dag a}  \o^\star [Y^j , Y^k; X^\dag_a] \, .  
\eea
In the above we have introduced $\k = \frac{2\pi}{k}$ and $\tilde{\Psi}^{\dag A} = \Psi^{\dag A}\g^2$.

\subsubsection{Case 1. Half Dirichlet}   
For the first case we shall assume that $Y$'s obey Dirichlet conditions i.e. $D_{\hat{\m}} Y = 0$. For the time being we solve this condition by setting $Y^i = 0$ which seems to be the simplest possibility but note that there may be other options. From the vanishing of the normal component of the super-current we have:
\bea
\label{ABJMeq1} 0&=&  \e_{ab} \tilde{\chi}^{\dag a} \g^2 \o D_2 X^b + \e_{ab}  \tilde{\chi}^{\dag a} \o \k [X^b, X^c; X^\dag_c] +  \e_{cd}  \tilde{\chi}^{\dag a} \o \k [X^c, X^d; X^\dag_a] \, , \\
\label{ABJMeq2}0&=&   \tilde{\xi}^{\dag i} \g^2 \o_{ai} D_2 X^a +   \tilde{\xi}^{\dag i} \o_{ai} \k [X^a, X^c;  X^\dag_c] \, ,  \\
\label{ABJMeq3}0&=&     \tilde{\chi}^{\dag a} \g^2 \o_{ai} D_2 Y^i  \, ,  \\
\label{ABJMeq4}0&=&   \e_{ij} \tilde{\xi}^{\dag i} \g^2 \o^\star D_2 Y^j \, ,  \\
\label{ABJMeq5}0&=&  \e_{ab} \tilde{\chi}^{\dag a} \g^{\hat{\m}} \o D_{\hat{\mu}} X^b    \, ,  \\
\label{ABJMeq6}0&=&    \tilde{\xi}^{\dag i} \g^{\hat{\m}} \o_{ai}  D_{\hat{\mu}} X^a  \, .
\eea
We are assuming that the $X$'s do not obey Dirichlet conditions (i.e.  $D_{\hat{\m}} X \neq 0$) and so in order to satisfy equation (\ref{ABJMeq5}) we thus require
\bea
0=  \tilde{\chi}^{\dag a} \g^{\hat{\m}} \o \,. 
\eea
We solve this by ansatz for the boundary condition on the fermions:  
\bea
\cP_+ \o  = \frac{1}{2} (1+\g^2 ) \o = 0 \ ,  &  \tilde{\chi}^{\dag a}  \cP_+ = 0 \, . 
\eea
Then we look at (\ref{ABJMeq3}) and demand $D_2 Y \neq 0$ which leads to
\bea
0 = \tilde{\chi}^{\dag a} \g^2 \o_{ai} = \tilde{\chi}^{\dag a}\cP_-  \g^2 \o_{ai} = - \tilde{\chi}^{\dag a} \g^2\cP_-  \o_{ai}
\eea
and see that we must have 
\bea
\cP_-  \o_{ai} = 0\, .
\eea
Then from (\ref{ABJMeq6}) we have 
\bea
0=\tilde{\xi}^{\dag i} \g^{\hat{\m}}\o_{ai}  = \tilde{\xi}^{\dag i} \g^{\hat{\m}} \cP_+ \o_{ai} = \tilde{\xi}^{\dag i} \cP_- \g^{\hat{\m}} \o_{ai}
\eea
and conclude 
\bea
 \tilde{\xi}^{\dag i } \cP_- = 0  \, . 
\eea
With these choices (\ref{ABJMeq4}) trivially solved. 

Then (\ref{ABJMeq2}) is solved by invoking a boundary condition for the boson
\bea
\label{BHeqn2}
D_2 X^a + \kappa  [X^a, X^c; X^\dag_c]  = 0 
\eea
and (\ref{ABJMeq1}) is further solved provided the boson also obeys 
\bea
-\e_{ab} D_2 X^b + \kappa  [X^c, X^b; X^\dag_c]  + \kappa \e_{cd}[ X^c, X^d;  X^\dag_a]  = 0 \,. 
\eea
At first these two conditions look strange, one does not expect two boundary conditions for a single field, however they are equivalent due to the identity 
\bea
\e_{cd}[ X^c, X^d;  X^\dag_a] = -2 \e_{ab}  [X^b, X^c; X^\dag_c]\,. 
\eea

In summary this solution has four bosons, $Y^i$ obeying Dirichlet boundary conditions and four bosons $X^a$ obeying Basu-Harvey type equations and the fermionic partners obey the appropriate corresponding projectors.    The preserved supersymmetry is generated by $\o_-$ and $\o_{ai +} $.  This is a curious feature, since $\o_-$ has only two real degrees of freedom whereas $\o_{ai}$ has four real degrees of freedom,  we have found an imbalance between left and right moving supersymmetries! 

\subsubsection{Spacetime Interpretation of preserved symmetry} 

The ABJM model with manifest ${\cal N}= 6$ supersymmetry at level $k$ is thought to describe membranes  in a $\mathbb{Z}_k$ orbifold background in which  $\mathbb{Z}_k$ acts by simultaneous rotation of the four complex planes in the transverse space.   At level $k=1,2$ this theory ought to be enhanced to the full ${\cal N}= 8$ supersymmetry.   In this section we investigate the details of this enhancement for the open membrane situation from both the spacetime perspective and from the world sheet boundary condition perspective.  

We first review the reasoning for ${\cal N}= 6$ supersymmetry from a space-time perspective.  As we have seen, the membrane breaks $SO(10,1)$ down to $SO(8) \times SO(1,2)$ and the supersymmetry parameter is restricted to obey
\bea
 \label{X-1} \Gamma^{012} \e = \e \,. 
\eea
To be explicit can decompose the gamma matrices as follows
\bea
\label{gammadec}
\Gamma^\m &=& \g^\m \otimes \bar{\Gamma}\, \quad \mu = 0,1,2  \\ 
\Gamma^{I+3} &=& 1 \otimes \Gamma^{I+1}\, \quad I = 0, \dots 7
\eea
and the supersymmetry parameter can be written as $\e = \e^{(2)}
\otimes \eta^{(8)} $ and then the projection (\ref{X-1}) implies that
$\eta^{(8)}$ is chiral.   It is convenient to work with a basis of
spinors $ \zeta^{( {\bf s} )}$ characterized by their  weight vector
${\bf s} = (s_1, s_2, s_3, s_4)$ where the entry $s_a$ takes values
$\pm \frac{1}{2}$ and is the eigenvalue of the appropriate generator of rotations
given by the product of the two gamma matrices defining the plane of rotation.  The chiral condition implies that there must be an even number of negative entries in the weight vector.

Now we look at the orbifold action 
\be
y^A \rightarrow e^{2\pi i / k} y^A \,
\ee
which has a corresponding action on $\eta^{(8)}$ of 
\be
\eta^{(8)} \rightarrow e^{2\pi i (s_1 + s_2 +s_3+ s_4) / k}\eta^{(8)} \,. 
\ee 
Demanding that the supersymmetry parameter is not projected out we must have that
\be
s_1 + s_2 + s_3 + s_4 = 0\, mod \,  k \,. 
\ee
For $k>2$ we can only solve this by $s_1 + s_2 + s_3 + s_4 = 0$ which means that the two weight vectors, $(+,+,+,+)$ and $(-,-,-,-)$, are projected out whence we find  ${\cal N}= 6$ supersymmetry.

Now we consider adding some five branes to the picture.  Since we can only add the five branes in a way that is compatible with the orbifold action, we have the projector conditions on the supersymmetry parameter 
\bea
 \label{X-2} \Gamma^{012} \e &=& \e \,, \\
   \label{X-3} \Gamma^{013456} \e &=& -\e \, .  
\eea 
Note that we have made a convenient choice of the relative orientation of the fivebrane.   Since we have broken the symmetry down to $SO(1,1) \times SO(8)$ we write  $\e = \e^{(2)}_+ \otimes \eta^{(8)}_+ \oplus \e^{(2)}_- \otimes \eta^{(8)}_- $ where $ \e^{(2)}_\pm$ are chiral $SO(1,1)$ spinors.   Using the decomposition of gamma matrices (\ref{gammadec}) we find that the spin weights that generate $\eta^{(8)}_+$ must obey
\bea
4S_1 S_2 \eta^{(8)}_+ = - \eta^{(8)}_+ \, , \quad 4S_1 S_2 \eta^{(8)}_+ = \eta^{(8)}_+ \,. 
\eea
Then we see that $\eta^{(8)}_+$ is generated by the four following weight vectors
\bea
(+,-,+,-), \, (+,-,-,+), \, (-,+,-,+), \, (-,+,+,-) \, ,  
\eea
whereas the left moving supersymmetry $\eta^{(8)}_-$ is generated by the two weights
\bea
(+,+,-,-), \, (-,-,+,+)\,.
\eea 
For levels $k=1,2$ we see that these left moving supersymmetries are augmented by the weight $(+,+,+,+)$ and $(-,-,-,-)$.   This discussion provides a space-time interpretation of the world sheet symmetries preserved by the above membrane boundary conditions.

\subsubsection{Enhancement of Supersymmetry} 
The preceding discussion indicates that we should expect supersymmetry enhancement in our boundary conditions for Chern Simons levels $k=1,2$.   It is thought that monopole operators are crucial to this process.   These local operators, which can best be thought of as creating a flux through a sphere surrounding their insertion point, are not gauge invariant.  In fact many different sorts of monopole operators can be built and they can be characterized by their non-abelian charges in terms of Young diagrams  (\cite{Klebanov:2009sg} provides a helpful review of this).  One important fact is that the minimum length of the rows in the tableaux are governed by the Chern Simons level $k$.  Thus for $k =1,2$ (and only these values) we have monopole operators 
\bea
(M^{(2)})_{p q}^{\hat{p} \hat{q}}\,   \quad (M^{(-2)})^{p q}_{\hat{p} \hat{q}}\,. 
\eea 
where we have explicitly indicated the $U(N)\times U(N)$ indices using lower case Roman indices $\{p,q\}$. 
Using these operators one can supplement the sixteen global symmetry currents, whose bosonic parts are given by  
\bea
J^A_{\mu B} = Tr\left( Y^A D_\mu Y^\dag_B - D_\mu Y^A Y^\dag_B \right) \ ,
\eea
with an extra six currents constructed with the monopoles
\bea
J^{AB}_\mu =    (M^{(-2)})^{p q}_{\hat{p} \hat{q}}  \left( Y^{A \hat{p} }_{p} D_\mu Y^{B\hat{q}}_{q}  - D_\mu Y^{A \hat{p} }_{p}  Y^{B\hat{q}}_{q}  \right) \, ,
\eea
and their six conjugates to give a complete set of twenty-eight currents of $SO(8)$. 
Furthermore such monopole operators are essential in being able to match operators of ABJM theory to KK modes given by symmetric traceless representations of $SO(8)$.   

Some recent proposal \cite{Gustavsson:2009pm, Kwon:2009ar} have been made to make the supersymmetry enhancement explicit by providing an extra set of ${\cal N}= 2$ supersymmetry transformations which supplement the ${\cal N}= 6$.  To do this we define some \textquoteleft non-ABJM\textquoteright \ fields 
\bea
W_A = M^{(2)} Z^\dag_A  = (M^{(2)})_{p q}^{\hat{p} \hat{q}} Z^{\dag q}_{A \hat{q}} 
\eea  
together with their fermionic partners $\Omega^A$ and their conjugates $\Omega^\dag_A$ and $W^{\dag A}$.  If we assume that the supercurrent corresponding to these extra supersymmetry transformations has the same form as that of the ${\cal N}= 6$ transformations i.e. 
\bea
\label{Jform}
J^\mu  = \Psi^{\dag A} \g^\m \delta \Psi_A + \delta \Psi^{\dag A} \g^\m \Psi_A   \ ,
\eea 
then from the supersymmetry rules given in \cite{Gustavsson:2009pm, Kwon:2009ar} we find a supercurrent given by 
\bea
J^2 = \tilde{\Psi}^{\dag A} \g^\nu \e D_\n W_A -  \tilde{\Psi}^{\dag A}  \e \lambda [W_A, Y^B; Y^\dag_B] + \frac{1}{3}   \tilde{\Psi}^{\dag A}  \e^\star \lambda [Y^B, Y^C; W^{\dag D}] \e_{ABCD} + h.c. \ , 
\eea
where $\lambda$ is a normalisation factor to be fixed shortly.    Important to these constructions is the fact that the monopole operators are covariantly constant.

Now we perform the same $SU(2) \times SU(2)$ decomposition we did above for the ${\cal N}=6$ supersymmetries letting $W_A = (u_a, v_i)$ and demand that $J^2=0$. This yields the following equations 
\bea
\label{Z-1} 0 &=&  \tilde{\chi}^{\dag a} \g^2 \e D_2 u_a -   \tilde{\chi}^{\dag a} \e \lambda [u_a , X^b ; X^\dag_b]  \, , \\
\label{Z-2} 0 &=&  \tilde{\xi}^{\dag i} \g^2 \e D_2 v_i -   \tilde{\xi}^{\dag i} \e \lambda [v_i , Y^j ; Y^\dag_j]  \, , \\
\label{Z-3} 0 &=& \tilde{ \chi}^{\dag a} \g^{\hat{\m}} \e D_{\hat{\mu}} u_a \, , \\
\label{Z-4}  0 &=& \tilde{\xi}^{\dag i} \g^{\hat{\m}} \e D_{\hat{\m}} v_i \, ,
\eea
together with two algebraic equations which involve three brackets with both $X$ and $Y$ fields.   

If we invoke the boundary conditions (Dirichlet for $Y$ and Basu-Harvey-Nahm-Neumann for $X$)  of the previous section and use the covariant constancy of the monopole operator we have that (\ref{Z-4}) is trivial and from (\ref{Z-2}) 
\bea
0 =  \tilde{\xi}^{\dag i} \g^2 \e  =  \tilde{\xi}^{\dag i} {\cal P}_+ \g^2 \e =    \tilde{\xi}^{\dag i}  \g^2 {\cal P}_+ \e \, .
\eea 
This implies must demand that the preserved supersymmetry parameter obeys $ {\cal P}_+ \e  = 0 $ i.e. that it is left moving.  This choice also solves (\ref{Z-3}). We can immediately see that this combines with the parameters $\o_-$ to restore the anticipated ${\cal N} = (4,4)$ supersymmetry we expect for a membrane ending on a five brane.   

All that remains is to solve equation (\ref{Z-1}), which we do by setting
\bea
\label{Z-5}
D_2 u_a + \lambda [u_a , X^b ; X^\dag_b] = 0\ ,
\eea
which, using the definition $u_a = M^{(2)} X^\dag_a$ and that the monopole operators are covariantly constant, we may write as
\bea
\label{Z-6}
M^{(2)}D_2 X^\dag_a + \lambda [M^{(2)} X^\dag_a , X^b ; X^\dag_b] = 0\,.
\eea
Using the Basu-Harvey equation (\ref{BHeqn2}), the condition (\ref{Z-6}) with the normalisations set so that $\l = \k$  represents a constraint on the monopole operator
\bea
 M^{(2)}[X^\dag_c , X^\dag_a; X^c] + [M^{(2)} X^\dag_a , X^b ; X^\dag_b] = 0\,. 
\eea

We remark that similar constraints involving monopole operators and three brackets were found necessary in the work of \cite{Gustavsson:2009pm} in order that the extra ${\cal N}=2$ supersymmetries closed and had an appropriate algebra with the ${\cal N}= 6$ supersymmetry.  

\subsubsection{Closure of Boundary Conditions}

Having examined the enhancement of supersymmetry we now return to our set of ${\cal N}=6$ boundary conditions.  An important question is whether our boundary conditions are closed under supersymmetry.   The supersymmetry variation of a boundary condition yields a new equation which needs to be satisfied.  Either this new equation will be a trivial consequence of the existing boundary conditions or it represents a new constraint which must be solved.  One may proceed in this way until we either have a closed set of boundary or an infinite number of non-trivial equations.  Given that we have remained supersymmetric in our derivation of the boundary conditions it is almost self evident that we expect the boundary conditions to close.  Nevertheless it is instructive to see this process in action.

In our case we need to understand the variation of our boundary conditions under the preserved supersymmetry $\o_-$ and $\o_{ai+}$.  We will begin by showing how the closure works for the $\o_{-}$ supersymmetry.  First we decompose the supersymmetry rules according to the $SU(2)\times SU(2)$ ansatz and implement the boundary condition on the resulting variation to find 
\bea
\delta X^a &=& i \e^{ab} \o^\star_- \chi_{b+} \\
\delta \chi_{a+} &=&  \e_{ab} \g^{\hat{\mu}} \o_- D_{\hat{\mu}} X^b\\
\delta \xi_{i-} &=&- \e_{ij} \o^\star_- D_2 Y^j \\
\delta A_2 &=& \k \left( \e^{ab} \o_{-}^\star \chi_{a+} X^\dag_b -   \e_{ab}  X^b \chi^{\dag a}_{+} \o_-  \right) \\
\delta \hat{A}_2 &=& \k \left( \e^{ab} \o_{-}^\star  X^\dag_b \chi_{a+}  -  \e_{ab} \chi^{\dag a}_{+}  X^b \o_-  \right)  \\
\delta Y^i &=& \delta \chi_{a-} =  \delta \xi_{i+} =\delta \hat{A}_{\pm} = \delta A_{\pm} = 0 
\eea
From these it is clear that the following boundary conditions are invariant
\bea
0 = Y^i = D_{\hat{\mu}} Y^i = \xi_{i+} = \chi_{a-}\, .
\eea
The Basu-Harvey equation, however, is not invariant.  In fact its variation results in  
\bea
\delta\left(D_2 X^a + \k [ X^a, X^c; X^\dag_c ] \right)  &=& D_2 \delta{X^a} + \kappa [\delta{X^a}, X^c; X^\dag_c]  -2 \kappa [\delta X^d, X^a; X^\dag_d] \\
&=& i \e^{ab} \o^\star_- \left(  D_2 \chi_{b+} + \k [\chi_{b+} , X^d; X^\dag_d]    \right) + 2 \k \e^{de} \o^\star_- [X^a, \chi_{e +} ; X^\dag_d] \, . \nonumber
\eea
This variation must also be set to zero and hence we conclude, after contraction with an extra epsilon symbol, 
\bea
D_2 \chi_{f+} - k \left( [\chi_{f+} , X^d; X^\dag_d] +2 [X^d, \chi_{d+} ; X^\dag_f] \right) = 0 
\eea
The equations of motion for the fermion field is given as
\bea
\gamma^\mu D_{\mu} \Psi_A = \kappa\left( [\Psi_{A}, Y^B; Y^\dag_{B}] +2 [Y^B, \Psi_B; Y^\dag_A] -2 \e_{ABCD} Y^B \Psi^{\dag C} Y^D \right)\ .
\eea
By continuity these equations should also be valid when restricted to the boundary.  We should only expect to be able to close the boundary conditions up to equations of motion (after all the supersymmetry algebra of the ABJM theory closes only on-shell).   Thus in evaluating the closure of boundary conditions we may use these equations, restricted to the boundary, evaluated with their boundary conditions imposed.  Firstly imposing the bosonic boundary conditions gives
\bea
0&=& \left. \left( \g^\mu D_\mu \chi_{a} - \kappa \left( [\chi_a , X^b ; X^\dag_b ] +2 [X^b, \chi_b ; X^\dag_a ] \right) \right) \right|_{\partial {\cal M}}\, , \\
0&=& \left. \left( \g^\mu D_\mu \xi_{i} - \kappa \left( [\xi_i , X^b ; X^\dag_b ] +2\e_{ij} \e_{ab} X^a \xi^{\dag j } X^b \right) \right)\right|_{\partial {\cal M}}\, .
\eea
Then we can apply projectors and invoke the fermionic boundary conditions to find in particular that
\bea
\label{chieq} 0&=& \left. \left( D_2 \chi_{a+} - \kappa \left([\chi_{a+} , X^b ; X^\dag_b ] +2 [X^b, \chi_{b+} ; X^\dag_a ]  \right)\right)\right|_{\partial {\cal M} } \ , \\ 
\label{xieq} 0&=&\left. \left( D_2 \xi_{i-} +  \kappa \left( [\xi_{i-} , X^b ; X^\dag_b ] +2\e_{ij} \e_{ab} X^a \xi^{\dag j }_- X^b \right) \right)\right|_{\partial {\cal M} }\, .
\eea
We now observe that the variation of the Basu-Harvey equation under the $\omega_-$ transformations vanishes due to the fermion equation of motion (\ref{chieq}).  One should not think of the variation of the Basu-Harvey equation producing any extra boundary conditions but rather being automatically zero as a consequence of the continuity of the equations of motion.   In a similar way one can calculate the variation of the Basu-Harvey equation under the $\omega_{ai+}$ supersymmetry.  One finds that the resultant variation is proportional to the $\xi_{i-}$ equation (\ref{xieq})\footnote{The calculation makes use of the identity (\ref{omegaid}).}.    This completes the proof of closure of the boundary conditions under the preserved supersymmetry.   

\subsubsection{Boundary Conditions and Classical Gauge Anomaly}

We have now established the correct form of the boundary conditions for the matter fields of the ABJM model but have not fixed boundary conditions on gauge fields.  Note that this is a significant difference to Yang-Mills theory where the supercurrent contains the field strength and so gives gauge field boundary conditions.   

The behaviour of the gauge field components  $A_2$ and $\hat{A}_2$ is actually tied algebraically to the boundary behaviour of the scalars through the covariant derivative term of the Basu-Harvey equation (\ref{BHeqn2}).   This is similar, and related to, the fact that the gauge field equation of motion ties current to flux in a Chern-Simons matter theory.    

 We must now establish suitable boundary conditions on the other components $A_{\hat{\mu}}$ and $\hat{A}_{\hat{\mu}}$. In what follows we find it convenient to work with the light-cone combinations defined as $A_\pm = A_0 \pm A_1$.   

A well known feature of Chern-Simons theories is that they are gauge invariant up to a boundary term. Therefore when the manifold has a boundary we find that there is a classical anomalous gauge transformation.  For the ABJM theory this is given by 
\bea
\delta_{gauge} {\cal L} = \frac{k}{4 \pi} \e^{\mu \nu \lambda} \partial_{\mu} Tr \left( \Lambda \partial_\nu A_\lambda -\hat{\Lambda} \partial_\nu \hat{A}_\lambda  \right)
\eea
A natural choice is then to chose boundary conditions on the gauge fields that eliminate this anomalous gauge variation.   A particularly appealing choice is to set the two gauge fields equal at the boundary\footnote{We thank E. Witten for suggesting these boundary conditions. }
\bea
\label{wbc}
\left. A_{\pm} \right|_{\partial \cM} = \left. \hat{A}_{\pm} \right|_{\partial \cM}\,.  
\eea
This boundary condition must be preserved by gauge transformations hence we require that the gauge transformation parameters are restricted by
\bea
\left. \Lambda \right|_{\partial \cM} = \left. \hat{\Lambda} \right|_{\partial \cM}\,.  
\eea
Then the gauge freedom that is preserved by this boundary condition is  the diagonal $U(N)$ of $U(N)\times U(N)$. This acts as gauge transformations with equal gauge parameters for both group factors.    This solution also has the virtue of preserving the parity operation in the following sense.  The three dimensional parity operator is given by a reflection in any one of the two spatial coordinates.  When there is a boundary these two operations are distinguishable, let us then consider for concreteness the parity operator given by $P: \{ x^0, x^1 , x^2 \} \rightarrow  \{ x^0, -x^1 , x^2 \}$.  Chern Simons theory is not parity invariant, however ABJM theory is, provided that parity acts on the gauge fields by additionally swapping $A$ with $\hat{A}$ so that:
\bea
P: \{A_0(x), A_1(x), A_2(x)\} \rightarrow   \{\hat{A}_0(x'), -\hat{A}_1(x'), \hat{A}_2(x')\} \, , \nn \\ 
P: \{\hat{A}_0(x), \hat{A}_1(x), \hat{A}_2(x)\} \rightarrow   \{A_0(x'), -A_1(x'), A_2(x')\} \, . 
\eea      
Thus, one can see that the boundary conditions for the gauge fields respect this operation.  Moreover, due to the boundary conditions on the scalars and fermions the supersymmetry variations of these gauge field boundary conditions is zero; they preserve supersymmetry.   

A further interesting feature of this choice of boundary condition for the gauge field is the covariant derivative acting on bifundametal matter becomes, on the boundary, in effect, the covariant derivative acting on matter in the adjoint representation of the surviving group i.e. :
\bea
D_{\hat{\mu}} Y^I |_{\partial \cM} = \left( \partial_{\hat{\mu}} Y^I   + i [ A_{\hat{\mu}}, Y^I] \right)  |_{\partial \cM}\,.  
\eea

A small point to mention is that because under the preserved supersymmetries $\delta A_2 \neq \delta \hat{A}_2$ one should not try to also impose that $A_2 = \hat{A}_2$.

There are, of course, other ways of removing this gauge anomaly for example by introducing extra fields on the boundary \cite{Chu:2009ms}, or more generally coupling to a boundary CFT.  In pure Chern-Simons theory we can choose boundary conditions on the gauge field that result in the famous derivation of the WZW model on the boundary (this approach has been looked at in  \cite{Berman:2009kj}).  In this context natural boundary conditions that also respect the above parity operation are to demand  $A_{+} = 0$ and $\hat{A}_{-}=0 $.

\subsubsection{Half Dirichlet with non-vanishing $Y$} 
For the case discussed above we have assumed that the Dirichlet condition $D_{\hat{\mu}}Y^i = 0$ is solved by setting $Y=0$.  One may ask are there other possibilities. For $Y\neq 0 $ the boundary conditions equations require additionally 
\bea
\label{ynon}
 0 &=& D_{\mu} Y^i = \partial_{\hat{\mu}} Y^i   + i [ A_{\hat{\mu}}, Y^i]  \, \nn \\
0 &=& \epsilon_{jk} [Y^j, Y^k ; X^{\dag}_a ] \nn \\
0 &=& [X^b, Y^i ;  Y^\dag_j] 
\eea
Note that we have adopted the $A=\hat{A}$ boundary condition for the gauge field described above.  This has allowed us to replace the bifundamental representation with an adjoint representation.  This has the implication that we no longer need to distinguish between hatted and non-hatted gauge indices on the matter fields.  This has the consequence that we can legitimately introduce a commutator of matter fields on the boundary. So for instance we can solve the above equations by demanding that for a constant $Y$
\bea
[X^a, Y^i] = [ X_a^\dag, Y^i] = [X^a , Y^\dag_i] = [Y^i, Y^j] =  [ A_{\hat{\mu}}, Y^i] =0 \,. 
\eea
That is the $Y$s need to be central in the remaining diagonal $U(N)$.   This allows the possibility of each of the $N$ membranes ending on different, parallel, fivebranes. This is achieved by decomposing $Y$ into basis elements of the Cartan subalgebra of $U(N)$. Then one chooses different values for each element. This is similar to the D3 case considered in   \cite{Gaiotto:2008sa}.  

Note that whilst these conditions are sufficient to solve (\ref{ynon}) they need not be necessary. A related question is the possibility of other branches to the moduli space of the ABJM model.

\subsubsection{A comment on anomalies}

The boundary conditions invoked on the fermionic fields imply that there are active chiral fermions on the boundary transforming in different R-symmetry representations.   One should therefore be concerned about possible anomalies.  For this case of a membrane ending on a fivebrane these anomalies are, in fact, canceled through two different inflow mechanisms. The tangent bundle anomaly is canceled as a result of the coupling of the two form potential on fivebrane to the self dual string.   The normal bundle anomaly is canceled by a contribution from the pullback of the Ganor-Motl-Intrilligator term on the fivebrane \cite{Ganor:1998ve,Intriligator:2000eq}. The details of this have been reported in \cite{bermanharvey}.  In the case of the membrane ending on the M9 brane described below, one would expect anomaly cancellation to occur along the lines of \cite{Horava:1995qa,Horava:1996ma} for the M-theory origin of the heterotic string.

\subsubsection{A comment on C-field backgrounds}

For a single membrane, the boundary conditions are altered by the presence of a constant C-field background.  This is because the membrane coupling becomes a surface term for a constant C-field. So the Dirichlet bosonic boundary condition (for the fivebrane) is modified to be 
\bea
\label{Cfieldbc}
\partial_2 X^A + \e^{\hat{\mu} \hat{\nu}} C^A_{\ BC} \partial_{\hat{\mu}} X^B \partial_{\hat{\nu}} X^C = 0 \, . 
\eea
This effect has been analysed in \cite{Bergshoeff:2000jn}.  Note the boundary condition (\ref{Cfieldbc}) does not simply mix Dirichlet with Neumann as would be the case for a string in a constant B-field background.  The preserved supersymmetries of this system are ``rotated" and the spacetime interpretation is where the membrane is no longer orthogonal to the fivebrane but instead has an angle $\theta$ to the fivebrane normal ($\theta$ is determined by the C-field) \cite{Michishita:2000hu,Youm:2000kr}.       

It is natural to repeat our analysis to include the effects of the C-field for the boundary conditions in the interacting theory.  A necessary first step is to determine how the C-field couples to the BL/ABJM theories.  The appropriate non-abelian ``pullback" has been considered in  \cite{Lambert:2009qw}. This is an interesting direction to pursue but the non-triviality of the non-abelian ``pullback" creates technical challenges which we leave for future work.

 \subsubsection{Case 2. No Dirichlet}
 Neither X nor Y obey Dirichlet conditions and so none of the equations are trivial.  To start we must have 
 \bea
0 &=& \e_{ab} \chi^{\dag a} \g^{\hat{\m}} \o D_{\hat{\m}} X^b \\
0 &=&  \e_{ij} \xi^{\dag i} \g^{\hat{\m}}\tilde{\o} D_{\hat{\m}} Y^j  \\
0 &=& \chi^{\dag a} \g^{\hat{\m}} \o_{ai} D_{\hat{\m}} Y^i  \\
0 &=&  \xi^{\dag i} \g^{\hat{\m}} \o_{ai} D_{\hat{\m}} X^a    
 \eea
 Up to overall chirality choice (orientation) these are solved by 
\bea
\chi^{\dag a}\cP_- =  \xi^{\dag i}\cP_- =  \cP_- \o = \cP_- \o_{ai} = 0
\eea
We now must solve all the remaining equations.  Similar to the case of BL one finds that 
\bea
D_2 X^A = D_2 Y^P = 0
\eea
and algebraic constraints expressed in the vanishing of three brackets between the scalars.  Again, these constraints have an interpretation in terms of the target space of the boundary string and it is interesting to explore the connection of this target space with the moduli space of the ABJM theory.  The space time interpretation of the supersymmetry enhancement is quite trivial, the six chiral supersymmetries of $SO(1,1)$ are extended to eight when the Chern-Simons level is $k=1,2$.   The gauge field boundary conditions described above are also applicable to this case.

\subsubsection{Case 3. All Dirichlet} 
We now move on to the case where all scalar fields obey Dirichlet boundary conditions and, for simplicity, we assume that they vanish. We can also demand that $D_2 X \neq 0$ and $D_2 Y \neq 0$. The only non trivial equations for supersymmetric boundary conditions that remain are 
\bea
0 &=& \e_{ab} \chi^{\dag a} \g^2 \o D_2 X^b \ , \\
0&=& \e_{ij} \xi^{\dag i} \g^2 \tilde{\o} D_2 Y^j\ ,\\
0 &=& \chi^{\dag a} \g^2 \o_{ai} D_2 Y^i\ , \\
0 &=& \xi^{\dag i} \g^2 \o_{ai} D_2 X^a \ .
\eea
These are solved by restricting the fermions as follows: 
\bea
\cP_+ \o = \cP_+ \o_{ai} \\
\chi^{\dag a}\cP_- =  \xi^{\dag i}\cP_-  = 0\,. 
\eea 

Although not all supersymmetry is manifest we believe that the interpretation of this case is the same as the corresponding situation in Bagger-Lambert theory described earlier.   Again the space time enhancement of supersymmetries is clear and in this case the closure of the boundary conditions is readily seen (and does not require use of the equations of motion).

\subsection*{Acknowledgments} 
DSB is partly supported by an STFC rolling grant and would like to thank DAMTP for continued hospitality. DCT is supported by an STFC studentship. Both DSB and DCT would like to thank Andrew Low for several helpful discussions on this work. The research of ES has been supported in part by
NSF Grant PHY-0555575. ES is also grateful to Cambridge-Mitchell Collaboration for financial support, and the Cambridge Centre for Theoretical Cosmology, DAMTP, for hospitality.

\section{Appendix}
 
\subsection{Conventions} 

\subsubsection{SO(2,1) Spinors}
Metric is given by $\eta_{\m\n} = \mathrm{diag}(-1,1,1)$ and the gamma matrices which obey $\{\g_\m \g_\n \} = 2 \eta_{\mu\nu}$ can be represented 
\bea
\g^0= \left(\begin{array}{cc} 0& 1\\ -1&0  \end{array} \right) = i \s_2\,,& \g^1= \left(\begin{array}{cc} 0& 1\\ 1&0  \end{array} \right) = \s_1 \, & \g^2= \left(\begin{array}{cc} 1& 0\\0&-1  \end{array} \right) =  \s_3\,. 
\eea
In addition to the Clifford algebra these also obey 
\bea
\g^\m \g^\n = \eta^{\m\n} + \e^{\m\n\l} \g_\l \, . 
\eea
   Note that these gamma matrices are defined with natural spinor index position $\g^{\m \, \b}_\a$. 
  Spinor indices can be raised and lowered with 
\be \e_{\a \b} = \left(\begin{array}{cc} 0& 1\\ -1&0  \end{array} \right)\,, \ee 
such that $\t^\a = \e^{\a \b}\t_\b$ and $\t_{\a} = \e^\b \e_{\b \a}$. With indices lowered the gamma matrices $\g^\m_{\a\b}$ are symmetric.  

 The natural index contraction is always NW to SE ($\searrow$) so that $\t \l = \t^\a \l_\a = \e^{\a\b} \t_\b \l_\a$  and we define $\t^\a \t_\a \equiv \t^2$.  In three dimensions we are able to form a contraction $\t \t^\ast$ which one can not do in four dimensions. For this reason when dealing with non-Majorana spinors we will find it convenient to avoid introducing over bars but to mark complex conjugation explicitly. For instance we have 
 \bea
 \o^\star \eta = \e^{\a \b} \o^\star_{\a} \eta_{\b}\ , \quad  & \o \eta = \e^{\a \b} \o_{\a} \eta_{\b}\,.
 \eea

\subsubsection{SO(6) Gamma Matrices}

The six gamma matrices which allow conversion between $SO(6)$ and $SU(4)$ can be realized as \cite{Terashima:2008sy}: 
\bea
\G^1 = \s_2\otimes 1\ ,  & \G^2 = - i \s_2\otimes \s_3 \ , &  \G^3 = i \s_2\otimes \s_1\ , \nn \\
\G^4 = -\s_1\otimes \s_2 \ , & \G^5 =  \s_3\otimes \s_2  \ ,&  \G^6 =  - i 1 \otimes \s_2  \ , \, 
\eea
 and obey
\bea
&&\G^i \tilde{\G}^{ j} + \G^j \tilde{\G}^{ i}= 2 \delta^{ij}\ .\\
&&\tilde{\Gamma}^{i AB} = -\frac{1}{2} \e^{ABCD} \G^i_{CD} = -(\G^{i}_{AB})^{\star}\ . 
\eea
The supersymmetry parameter can be written as $\o_{AB} \equiv \e^i \G^i_{AB}$ where $\e^i$ are six, two-component Majorana spinors.  

If we define complex (i.e. non Majorana spinors) as
\bea
\l_1 = \e^6+ i\e^5\ , & \l_2 = - \e^3 + i \e^4 \ , & \l_3= \e^2  +i \e^1\ ,
\eea
we can explicitly write 
\bea
\o_{AB} = \left(\begin{array}{cccc}
0& -\l_1 & -\l_3 & -\bar{\l}_2 \\
\l_1 &0 & - \l_2 & \bar{\l}_3\\
\l_3 & \l_2 & 0 & - \bar{\l}_1\\
\bar{\l}_2 & -\bar{\l}_3 & \bar{\l}_1 &0 
\end{array}\right)\ . 
\eea
The case where $\l_2=\l_3=0$ reduces to the supersymmetry rules  found in the $\cN = 2 $ superspace formulation \cite{Benna:2008zy}.  For example the scalar field transformation becomes
\bea
\de Y^1 = - i\bar{\l} \p_2 \ ,   &  \de Y^2 =  i\bar{\l} \p_1 \ , \nn \\
\de Y^3 = - i \l \p_4\ ,  & \de Y^4 = i \l \p_3 \ .
\eea
Notice that the manifest $SU(4)$ structure is broken down to $SU(2) \times SU(2)$.  In the $\cN = 2$ superspace formalism of \cite{Benna:2008zy}, we have two sets of two chiral superfield $\cZ^a$ and $\cW_a$ whose bosonic components $Z^a$ and $W_a$  are given in terms of the transverse coordinates as
\bea
Z^1 = X^1 + i X^5 \ ,  & W_1 = X^{\dag 3} + i X^{\dag 7}\ , \nn \\
Z^2 = X^2 + i X^6 \ , & W_2 = X^{\dag 4} + i X^{\dag 8}\ , 
\eea
and $SU(4)$ combination of these fields is $Y^A= \{ Z^1 ,Z^2, W^{\dag 1} , W^{\dag 2}  \}$.

\bibliographystyle{alpha}

\end{document}